\documentclass[11pt,a4paper]{article}
\usepackage[hyperref]{styles/acl2019}
\usepackage{times}
\usepackage{latexsym}
\usepackage{graphicx}
\usepackage{adjustbox}
\usepackage{multirow}
\usepackage{amsmath}
\usepackage{booktabs}
\usepackage{enumitem}
\usepackage{url}
\usepackage{footmisc}

\newcommand{\systemfull}[1]{MediQA Task}
\newcommand{\system}[1]{MediQA Task}

\newcommand{\remindersj}[1]{}

\usepackage{array}
\newcolumntype{L}[1]{>{\raggedright\let\newline\\\arraybackslash\hspace{0pt}}m{#1}}
\newcolumntype{C}[1]{>{\centering\let\newline\\\arraybackslash\hspace{0pt}}m{#1}}
\newcolumntype{R}[1]{>{\raggedleft\let\newline\\\arraybackslash\hspace{0pt}}m{#1}}

\aclfinalcopy 

\title{Pentagon at MEDIQA 2019: Multi-task Learning for Filtering and \\ Re-ranking Answers using Language Inference and Question Entailment}
\author{Hemant Pugaliya\thanks{ * Equal contribution, randomly sorted. Karan and Shefali took ownership of the NLI module while Sheetal and Prashant worked on the RQE module. Hemant researched and implemented the Question-Answering system including baseline and multi-task learning. Sheetal and Hemant worked on scraping data from icliniq. Karan and Prashant helped with integration of NLI and RQE module respectively into the multi-task system.},\quad Karan Saxena\footnotemark[1],\quad Shefali Garg\footnotemark[1],\quad Sheetal Shalini\footnotemark[1],\\\textbf{Prashant Gupta\footnotemark[1],\quad Eric Nyberg,\quad Teruko Mitamura} \\ \\ \textit{\{hpugaliy, karansax, shefalig, sshalini, prashang, ehn, teruko\}@cs.cmu.edu} \\ \\ Language Technologies Institute (LTI), \\ Carnegie Mellon University (CMU) }

\begin{document}
\maketitle
\begin{abstract}
Parallel deep learning architectures like fine-tuned BERT  and  MT-DNN, have quickly become the state of the art, bypassing previous deep and shallow learning methods by a large margin.
More recently, pre-trained models from large related datasets have been able to perform well on many downstream tasks by just fine-tuning on domain-specific datasets (similar to transfer learning).\\
However, using powerful models on non-trivial tasks, such as ranking and large document classification, still remains a challenge due to input size limitations\footnote{https://github.com/google-research/bert/issues/27} of parallel architecture and extremely small datasets (insufficient for fine-tuning).\\
In this work, we introduce an end-to-end system, trained in a multi-task setting, to filter and re-rank answers in the medical domain. We use task-specific pre-trained models as deep feature extractors. Our model achieves the highest Spearman's Rho and Mean Reciprocal Rank of 0.338 and 0.9622 respectively, on the ACL-BioNLP workshop MediQA Question Answering shared-task.
\end{abstract}

\section{Introduction}
In this work, we study the problem of re-ranking and filtering in medical domain Information Retrieval (IR) systems. Historically, re-ranking is generally treated as a `Learning to Rank' problem while filtering is posed as a `Binary Classification' problem. Traditional methods have used handcrafted features to train such systems. However, recently deep learning methods have gained popularity in the Information retrieval (IR) domain \cite{mitra2017neural}.

The ACL-BioNLP workshop MediQA shared task \cite{MEDIQA2019} aims to develop relevant techniques for inference and entailment in medical domain to improve domain specific IR and QA systems. The challenge consists of three tasks which are evaluated separately.

The first task is the Natural Language Inference (NLI) task which focuses on determining whether a natural language hypothesis can be inferred from a natural language premise. The second task is to recognize question entailment (RQE) between a pair of questions. The third task is to filter and improve the ranking of automatically retrieved answers.

For the NLI and RQE tasks, we use transfer learning on prevalent pre-trained models like BERT \cite{devlin2018bert} and MT-DNN \cite{liu2019mt-dnn}. These models play a pivotal role to gain deeper semantic understanding of the content for the final task (filtering and re-ranking) of the challenge \cite{2018arXiv180902922D}. Besides using usual techniques for candidate answer selection and re-ranking, we use features obtained from NLI and RQE models. We majorly concentrate on the novel multi-task approach in this paper. We also succinctly describe our NLI and RQE models and their performance on the final leaderboard.




\section{Related Work}
Past research demonstrates a simple architecture for filtering and re-ranking, where the system returns the best answer based on the Information Retrieval and Question Entailment Scores [ from a corpus of FAQs scraped from medical websites MediQUAD \cite{2019arXiv190108079B} ]. This system outperformed all the systems participating in the TREC Medical LiveQA’17 challenges. \cite{harabagiu-hickl-2006-methods} successfully shows the use of Natural Language Inference (NLI) in passage retrieval, answer selection and answer re-ranking to advance open-domain question answering. \cite{semanticrelated} shows effective use of UMLS \cite{UMLS}, a Unified Medical Language System to asses passage relevancy through semantic relatedness. All these methods work well independently, but to the best of our knowledge, there hasn't been much work in using NLI and RQE systems in tandem for the tasks of filtering and re-ranking. 

As noted in \cite{romanov2018lessons}, the task of Natural Language Inference is not domain agnostic, and thus is not able to transfer well to other domains. The authors use a gradient boosting classifier \cite{mason2000boosting} with a variety of hand crafted features for baselines. They then use Infersent \cite{conneau2017supervised} as a sentence encoder. The paper also reports results on the ESIM Model \cite{Chen_2017} but with no visible improvements. They also discuss transfer learning and external knowledge based methods. 

Most traditional approaches to Question Entailment use bag-of-word pair classifiers \cite{tungdetermining} using only lexical similarity. However, in the recent past, neural models \cite{mishradeep} have been employed to determine entailment between questions incorporating their semantic similarity as well. These techniques work by generating word-level representations for both the questions, which are then combined into independent question representations by passing it through a recurrent cell like Bi-LSTM \cite{liu2016learning}. However, current state-of-the-art methods like BERT \cite{devlin2018bert} and MT-DNN \cite{huang2013multi} learn a joint embedding of the two questions, which is then used for classification.

 \cite{abacha2016recognizing} implemented the SVM, Logistic Regression, Naive Bayes and J48 models as baselines for Question Entailment task. They use a set of handcrafted lexical features, like word overlap and bigram similarity, and semantic features like number of medical entities (problems, treatments, tests) using a CRF classifier trained on i2b2 \cite{uzuner20112010} and NCBI corpus \cite{dougan2014ncbi}.

\section{Dataset \& Evaluation}
The dataset for re-ranking and filtering has been provided by the MediQA Shared task \cite{MEDIQA2019} in ACL-BioNLP 2019 workshop. 
It consists of medical questions and their associated answers retrieved by CHiQA \footnote{https://chiqa.nlm.nih.gov/}. The training dataset consists of 208 questions while the validation and test datasets have 25 and 150 questions respectivley.
Each question has upto 10 candidate answers, with each answer having the following attributes : 
\begin{enumerate}
\item SystemRank: It corresponds to CHiQA's rank. 
\item ReferenceRank: It corresponds to the correct rank. 
\item ReferenceScore: This is an additional score that is provided only in the training and validation sets, which corresponds to the manual judgment/rating of the answer [4: Excellent, 3: Correct but Incomplete, 2: Related, 1: Incorrect].
\end{enumerate}
For the answer classification task, answers with scores 1 and 2 are considered as incorrect (label 0), and answers with scores 3 and 4 are considered as correct (label 1).
The evaluation metrics for filtering task is Accuracy and Precision while metrics for re-ranking task is Mean Reciprocal Rank (MRR) and Spearman's Rank Correlation Coefficient.

To train the Natural Language Inference and Question  Entailment module of our system we again use the data from MediQA shared task \cite{MEDIQA2019}. 

For Natural Language Inference (NLI), we use MedNLI \cite{romanov2018lessons} dataset. It is a dataset for natural language inference in clinical domain that is analogous to SNLI. It includes 15,473 annotated clinical sentence pairs. For our model, we create a training set of 14,050 pairs and a held out validation set of 1,423 pairs. The evaluation metric for NLI is accuracy.

The dataset used for Question Entailment (RQE) consists of paired customer health questions (CHQ) and Frequently Asked Questions (FAQ) \cite{RQE:AMIA16}. We are provided labels for whether FAQ entails CHQ or not. The RQE training dataset consists of 8,588 medical question pairs. The validation set comprises of 302 pairs. The evaluation metric used for RQE is accuracy.

We also augment the data from a popular medical expert answering website called \footnote{https://www.icliniq.com/qa/medical-conditions}{icliniq}. It is a forum where users can delineate their medical issues, which are then paraphrased as short queries by medical experts. The user queries are treated as CHQs whereas the paraphrased queries are treated as FAQs. We extract 9,958 positive examples and generate an equal number of negative examples by random sampling. The average CHQ length is 180 tokens whereas the average FAQ length is 11 tokens. In addition, the expert answers are used to augment the MediQUAD corpus \cite{2019arXiv190108079B}. 





\section{Approach/System Overview} \label{sec:approach}
We use pretrained RQE and NLI modules as feature extractors to compute best entailed questions and best candidate answers in our proposed pipeline.

\subsection{Pretraining NLI and RQE modules} \label{sec:nli-rqe}
Both the NLI and RQE modules use MediQA shared task \cite{MEDIQA2019}  for training (fine-tuning) and computing the inference and entailment scores. For both the tasks, we use the following approaches to preprocess the datasets:
\begin{enumerate}
    \item Replacing medical terms with their preferred UMLS name. We augment the terms like \textit{Heart attack} in the sentence with \textit{Myocardial infarction} extracted from UMLS.
    \item Expanding abbreviations for medical terms in order to normalize the data. The list of medical abbreviations is scraped from Wikipedia. Since this list of abbreviations also contains full forms of stop words like ``IS", ``BE", we manually curate the list to contain only the relevant acronyms.
\end{enumerate}

For fine-tuning the NLI and RQE modules, we use the dataset for NLI and RQE tasks of MediQA shared task \cite{MEDIQA2019} respectivley.  We also augment the RQE dataset with data from icliniq during fine-tuning.

\subsection{Preprocessing}
A lot of answers have spurious trailing lines about FAQs being updated. Any trailing sentences in the answers having ``Updated by:" are removed. A co-reference resolution is run on each answer using Stanford CoreNLP \cite{manning-EtAl:2014:P14-5} and all the entity-mentions are replaced with their corresponding names.

\subsection{Using RQE module} \label{rqe_module}
\begin{figure*}[h]
\centering
  \includegraphics[scale=0.5]{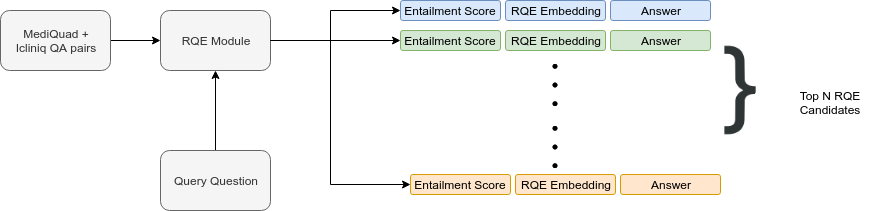}
  \caption{Finding entailed questions from MediQuad and Icliniq QA pairs, for a particular query. We get an entailment score, RQE entailment embedding, and answer for each QA pair in MediQuad and Iclinq data. We then pick up the top N entailed questions.}
  \label{fig:rqe}
\end{figure*}

For each question in the training  set  we  get  upto N entailing questions (along with their scores and embeddings) and answers with a threshold T for confidence using RQE module.
We use this system both in the baseline and the multi-task learning system. The complete process is highlighted in Figure \ref{fig:rqe}.

\subsection{Baseline: Feature-Engineered System} \label{feature}
 We develop a feature-engineered system as a baseline. This system uses the following features:
 \begin{enumerate}
    \item Answer Source (One-hot)
    \item Answer Length In Sentences
    \item ChiQA Rank
    \item Bag of Words(BoW) TF-IDF scores of Candidate Answer (trained on MediQUAD)
    \item Bag of Words (BoW) TF-IDF scores of 1-best Entailed answer (trained on MediQUAD)
    \item N-best RQE Scores
    \item N-best RQE embeddings
    \item N-best Average NLI Scores
\end{enumerate}
Average NLI score between the candidate answer `s' containing `S' sentences and entailed answer `p' containing `P' sentences is defined as:
\begin{center}
    $ANLI(s,p) = \frac{\sum_{S}(\max_{P}(NLI(S,P) ))}{|S|}$ \label{eq:ANLI}

\end{center}
where $|S|$ symbolizes the total number of sentences in candidate answer.\\
For a given confidence threshold T, if N candidates are not obtained from RQE model we set the corresponding features to 0. 

We train the system using the above features with Logistic regression for filtering and use the scores to rank the answers. We also train a system with same features using SVM-rank \cite{Joachims:2006:TLS:1150402.1150429} to improve our ranking metrics.  All the results have been discussed in Section \ref{results}.

\subsection{Jointly Learning to Filter and Re-rank} \label{joint}
 Multitask learning is defined as, ``A learning paradigm in machine learning whose aim is to leverage useful information contained in multiple related tasks to help improve the generalization performance of all the tasks". \cite{2017arXiv170708114Z}  As the tasks of both filtering and re-ranking are highly related and can benefit from shared feature space, we propose a multi-task learning based system to both rank and filter our candidate answers.

In this system we use the MT-DNN \cite{liu2019mt-dnn} based models developed for NLI and RQE (described in Sectition \ref{sec:nli-rqe}) as feature extractors. The embedding generated for classification from both the models is used as features. In addition we also use the scores from RQE models to get RQE candidates from MediQUAD \cite{2019arXiv190108079B} corpus. Going forward we refer to these features as embeddings.

Our initial step is the same as our baseline system and is summarized in Section \ref{rqe_module}. For each candidate answer in training set and the retrieved entailed answer we obtain the following embeddings:
\begin{enumerate}
    \item \textbf{NLI Embedding}: If an entailing question's answer A has \textit{a} sentences and candidate answers C have \textit{c} sentences, then a tensor of $$a*c*768$$ is extracted to make an embedding matrix using the NLI module.
    Each sentence in entailing Answer A is combined with every sentence of candidate answer C and passed to the MT-DNN NLI model to build this tensor. We then run a convolution encoder on this matrix to obtain an NLI embedding. The final layer is an average pooling layer which averages each of the four quadrants of 256 channel feature map and concatenates them to obtain an NLI embedding of size 1024. This step is necessary to convert varied size (due to varying \textit{a} and \textit{c} above) feature maps to a single embedding of size 1024.
\begin{figure*}[h]
\centering
  \includegraphics[scale=0.5]{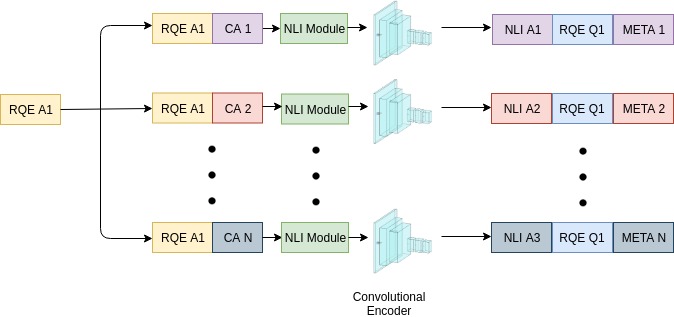}
  \caption{Creating NLI embedding for each RQE Answer A1 by concatenating it with every candidate answer (CA) and passing it through the NLI module and convolutional encoder. This NLI embedding is then concatenated with the corresponding RQE Question (Q1) embedding and Metadata embedding to obtain the joint embedding.}
  \label{fig:joint_cnn}
\end{figure*}

\item \textbf{RQE Embedding}: This is the embedding obtained from the RQE model while searching for the entailed questions.
 
 \item \textbf{Metadata Embedding}: This embedding encodes metadata features for the pair. We encode the candidate answer source (one-hot), the entailed answer source (one-hot), candidate answer length, entailed answer length, candidate answer system rank, and TF-IDF scores of 2000 words (trained on MediQUAD) for the candidate answer.

\end{enumerate}

 We concatenate the above embeddings for each candidate answer (referred to as joint embedding going forward). For a given entailed answer, one joint embedding is obtained for each candidate answer. The entire process of converting a single entailing answer to a set of joint embeddings for candidates is summarized in Figure \ref{fig:joint_cnn}.   
 
Using the joint embeddings obtained above we train two binary classifiers, which are fully connected neural networks, as follows:
 
 \begin{enumerate}
 \item \textbf{Filtering classifier}: This classifier takes in the joint embedding for a single candidate answer and classifies it as relevant or irrelevant.
 
\item \textbf{Pairwise ranking classifier}: This classifier takes in the joint embedding of two candidate answers and classifies if the first candidate ranks higher or not.
\end{enumerate}

Architecture details are provided in Appendix \ref{sec:appendix}.
 
\subsection{Training and Inference details}
For a given confidence threshold T, if less than N questions are obtained from the RQE model, we only use the questions which satisfy the threshold. In case no entailed question is returned, we use the top entailed question despite confidence being below the threshold. Unlike the baseline, here joint embedding is extracted separately for each entailed answer. Hence this allows for having different number of entailed answers for each question.

For training we consider each question and candidate answers as a batch. We define the final training loss as follows:
\begin{multline}
L_{total} = \sum_{N}(\sum_{c}L_{filter}(c) \\  + \alpha \sum_{p \in c^2 -1 pairs}L_{pair}(p))
\end{multline}
where \textit{N} is the number of RQE candidates we have for this question, \textit{c} is the number of candidate answers, $L_{filter}$ is the loss obtained from filtering classifiers, $L_{pair}$ is the loss obtained from pairwise classifier. We use Cross-Entropy Loss for both $L_{filter}$ and $L_{pair}$. Here we use $\alpha = 2$ to focus more on the re-ranking task as it is considered tougher than filtering. To augment the training data we use higher-ranked candidate answers as entailed answers to create training instances with lower-ranked candidate answers. 
While inference we use the ensemble from different RQE candidates to decide upon filtering and pairwise ranking, by summing the scores from candidates.
\section{Experiments and Results} \label{results}
For this task we perform multiple experiments on feature-engineered system in Section \ref{feature} to asses the usefulness of the designed features. These experiments later help us incorporate these features into Metadata Embedding defined in Section \ref{joint}.

Firstly we run the experiments on Metedata features, BoW, Coarse-grained RQE and NLI scores. The results are shown in Table \ref{tab:task3-coarse}. We later incorporate the RQE embeddings from RQE system and the results are shown in Table \ref{tab:task3-fine}. Here we evaluate the system with different number of RQE candidates at different threshold settings. Previous experiments were conducted on the filtering task only .  For ranking task we train SVM-Rank \cite{Joachims:2006:TLS:1150402.1150429} based systems to learn pair-wise ranking, using the same features as the filtering task.  Experiments with SVM-Rank \cite{Joachims:2006:TLS:1150402.1150429} were performed with N=3 RQE candidates and the results are shown in Table \ref{tab:task3-rank}.

Moving to jointly learning system introduced in Section \ref{joint}, we train it with different parameter settings. Due to lack of resources, we could evaluate only a few hyperparameter settings where N is the most number of RQE candidates considered while training and T is the the threshold for retrieving the candidates. In addition we also evaluate the results with augmented datasets from icliniq. We share the results on validation data in Table \ref{tab:task3-val} and results on test set in Table \ref{tab:task3_test_results}.

\begin{table}[]
\centering
\begin{tabular}{|p{2.7cm}|c|p{1.75cm}|}
\hline
\textbf{Metrics} & \textbf{Accuracy} & \textbf{Spearman’s Rho} \\ \hline
Metadata & 50.12 & 0.091 \\ \hline
Metadata + BoW + RQE Scores & 61.23 & 0.125 \\ \hline
Metadata + BoW + RQE scores + Avg NLI & 62.17 & 0.127 \\ \hline
\end{tabular}
\caption{Results with features except RQE embeddings using Logistic Regression }
\label{tab:task3-coarse}
\end{table}

\begin{table}[]
\centering
\begin{tabular}{|c|c|c|c|}
\hline
 & \multicolumn{3}{l|}{\textbf{No. of RQE Candidates}} \\ \hline
\textbf{RQE Threshold} & \textbf{N=1} & \textbf{N=3} & \textbf{N=5} \\ \hline
\textbf{No Threshold} & 63.12 & 67.17 & 64.92 \\ \hline
\textbf{T=0.9} & 63.21 & 65.12 & 63.9 \\ \hline
\textbf{T=0.7} & 64.91 & \textbf{69.67} & \textbf{66.123} \\ \hline
\textbf{T=0.5} & \textbf{65.18} & 68.96 & 65.031 \\ \hline
\end{tabular}
\caption{Accuracy obtained on including  RQE embeddings.}
\label{tab:task3-fine}
\end{table}

\begin{table}[]
\centering
\begin{tabular}{|c|c|}
\hline
\textbf{RQE Threshold} & \textbf{Coverage} \\ \hline
T=0.5 & 186/208 \\ \hline
T=0.7 & 175/208 \\ \hline
T=0.9 & 150/208 \\ \hline
\end{tabular}
\caption{Coverage of Validation set based on RQE threshold for Task 3.}
\label{tab:task3-coverage}
\end{table}

\begin{table}[]
\centering
\begin{tabular}{|l|l|}
\hline
\textbf{System} & \textbf{Spearman's Rho} \\ \hline
LR based filtering & 0.2327 \\ \hline
Rank-SVM(T=0.9) & 0.2627 \\ \hline
Rank-SVM(T=0.7) & \textbf{0.2972} \\ \hline
Rank-SVM(T=0.5) & 0.2812 \\ \hline
\end{tabular}
\caption{Rank-SVM results with Fine-grained features for N=3 candidates with different threshold levels for Task 3.}
\label{tab:task3-rank}
\end{table}

\begin{table*}[t]
\centering
\begin{adjustbox}{width=1\textwidth}
\begin{tabular}{|c|c|c|c|c|}
\hline
\textbf{Hyperparameters}                            & \textbf{Accuracy} & \textbf{Spearman's Rho} & \textbf{MRR} & \textbf{Precision} \\ \hline
N=3, T=0.7, Corpus = Mediquad + Icliniq & \textbf{0.765}             & 0.338                   & \textbf{0.962}        & \textbf{0.776}              \\ \hline
N=3 , T=0.7 , Corpus = Mediquad                     & 0.733             & \textbf{0.354}                   & 0.955        & 0.741              \\ \hline
N=5 , T=0.7, Corpus = Mediquad                      & 0.7               & 0.317                   & 0.97         & 0.709              \\ \hline
\end{tabular}
\end{adjustbox}
\caption{Multi-Task learning results with different parameter settings on Test data for Task 3.}
\label{tab:task3_test_results}
\end{table*}

\begin{table*}[t]
\centering
\begin{tabular}{|l|l|l|}
\hline
\textbf{Hyperparameters} & \textbf{Accuracy} & \textbf{Spearman's Rho} \\ \hline
N=3, T=0.7, Corpus = Mediquad + Icliniq & \textbf{78.12} & 0.351 \\ \hline
N=3 , T=0.7 , Corpus = Mediquad & 76.1 & \textbf{0.372} \\ \hline
N=5 , T=0.7, Corpus = Mediquad & 71.1 & 0.331 \\ \hline
\end{tabular}
\caption{Multi-Task learning results with different parameter settings on Validation data for Task 3.}
\label{tab:task3-val}
\end{table*}
\section{Discussion}

We design the experiments to see if the answering and re-ranking tasks can be improved upon using RQE and NLI tasks. This hypothesis was proved by seeing the improved performance on including RQE and NLI features in Section \ref{feature} as shown in Table \ref{tab:task3-coarse}. Moreover we see that on including RQE embeddings we get the performance boost as seen in Table \ref{tab:task3-fine}.

Another question which we can ask ourselves is how many entailed answers are good enough for performing filtering and re-ranking and how confident do we need to be about the entailment to consider a candidate.  Experimental results shown in Table \ref{tab:task3-fine} show that we can't take too high number of candidates as well as the threshold can't be too high. We see in Table \ref{tab:task3-coverage} that if we take too high threshold for entailment, we might not find an entailing answer altogether. Hence going forward for all experiments we have taken threshold as 0.7 and number of candidates as 3.

While the feature sets discussed in the above experiments perform well in filtering tasks, they do not do well when their re-ranking is done based on their filtering scores. In further experiments we train a specialized ranking system using SVM-Rank \cite{Joachims:2006:TLS:1150402.1150429} and the results are shared in Table \ref{tab:task3-rank}. We see that the same exact feature set could learn well to re-rank when trained with specialized algorithm. Improved results in Table \ref{tab:task3-fine} and Table \ref{tab:task3-rank} by learning on the same feature set but using different algorithms motivated us to design our approach in Section \ref{joint} which would learn a joint high-dimensional feature space for both the tasks.   

Experiments on multi-task learning clearly show that this technique is superior to feature-engineered approach in both re-ranking and filtering. We attribute this increase in performance to mainly two factors: Firstly the multi-task setting allows it to learn more generalized features. Secondly, inclusion of high-dimensional NLI features in the architecture which was previously not possible with feature-engineered approach. However the computationally expensive nature of this approach did not let us experiment with many hyperparameter settings. The results on Validation data and Test data are shown in Table \ref{tab:task3-val} and Table \ref{tab:task3_test_results} respectively.

From the results in Table \ref{tab:task3-val} and Table \ref{tab:task3_test_results} we see that it reinforces our analysis done about the candidate and threshold settings based on Table \ref{tab:task3-coverage}. We also see that adding additional data from Icliniq improves the accuracy but decreases the Spearman's Rho. This can be attributed to the language style difference between ICliniq and MediQUAD \cite{2019arXiv190108079B}. As re-ranking is a tougher task, it's performance takes a hit while the accuracy does improve owing to better RQE coverage.  

\section{Shared Task Performance}
To evaluate our performance on the test sets, we submitted our NLI, RQE and Re-ranking \& Filtering model independently on the shared task leader-board. For Task 1, i.e. the NLI task, we achieved an accuracy of 85.7 on the test set. 
For Task 2, i.e. the RQE task, we observed that the test set varied greatly as compared to the training set, leading to poor results on test dataset. To account for this difference, we discarded the training data and trained our model only on the validation and augmented data. This model gave us an accuracy of 67.1 on the test set.
The best model for both the tasks is the ensemble of Infersent \cite{conneau2017supervised}, BERT fine-tuned (last 4 layers) \cite{devlin2018bert} and MT-DNN \cite{huang2013multi}. For both  
For Task 3, i.e. the re-ranking and filtering task, the results are shown in Table \ref{tab:task3_test_results}.

In the NLI task , our system ranked 7\textsuperscript{th} (out of 17), showing an improvement of 20\% over the task baseline. In the RQE task, our system ranked  4\textsuperscript{th} (out of 12), showing an improvement of 24\% over the task baseline. In the Question Answering Task, our system ranked 3\textsuperscript{rd} (out of 10) in filtering metrics (both Accuracy and Precision) while it ranked 1\textsuperscript{st} (out of 10) in the ranking metrics (both Mean Reciprocal Rank and Spearman\'s Rho). The system performs significantly better than others in ranking metrics, showing an improvement of 2.6\% and 42\% in Mean Reciprocal Rank and Spearman's Rho respectively over the next best scores from the participating teams.  Interestingly, our system is the only participating system which outperforms the baseline (ChiQA provided answers) on Spearman's Rho. However, this is not surprising as we take ChiQA rank as one of our input features.

\section{Error Analysis}

\begin{figure}[h!]
  \includegraphics[scale=0.49]{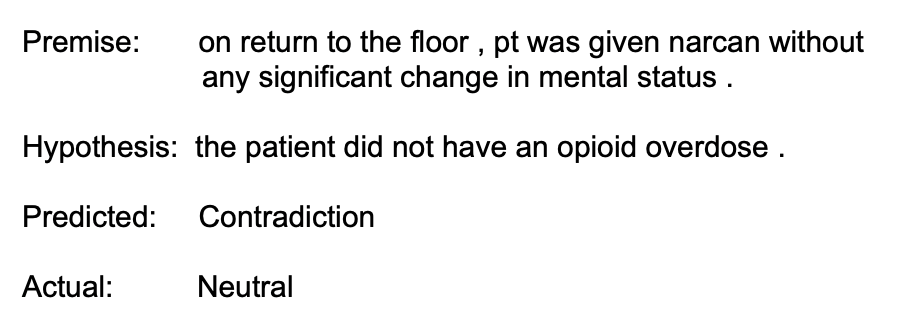}
  \caption{NLI model Incorrectly Predicting Contradiction on Test Set}
  \label{fig:error_neg_nli}
\end{figure}

In case of NLI, we observe that the model generally fails in two major settings explained below. \\
Since most of our training data has negation words like `do not', `not' etc for the contradicting hypothesis, the model assigns the label as contradiction whenever it sees a confusing example with negation term as shown in Figure \ref{fig:error_neg_nli}.

\begin{figure}[h!]
  \includegraphics[scale=0.42]{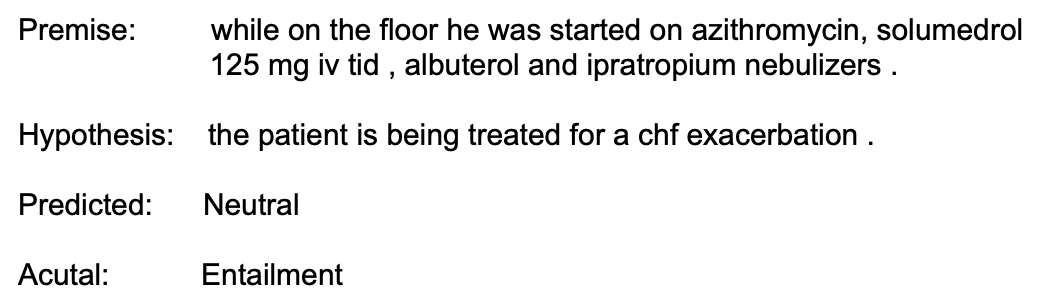}
  \caption{NLI model Incorrectly Predicting Neutral on Test Set}
  \label{fig:nli_error}
\end{figure}

The model also fails while trying to differentiate between statements that are neutral versus those that entail each other. The model generally relies on lexical overlap between the hypothesis and the premise, and in cases, when it is unable to find one, falls back to assigning the label as neutral as shown in Figure \ref{fig:nli_error}.

For the RQE task, we observe that our model labels the CHQ-FAQ pairs as entailment when they have a high lexical overlap of the medical entities and not entailment otherwise. We confirm this with some examples from the RQE test set. 

\begin{figure}[h!]
\centering
  \includegraphics[scale=0.49]{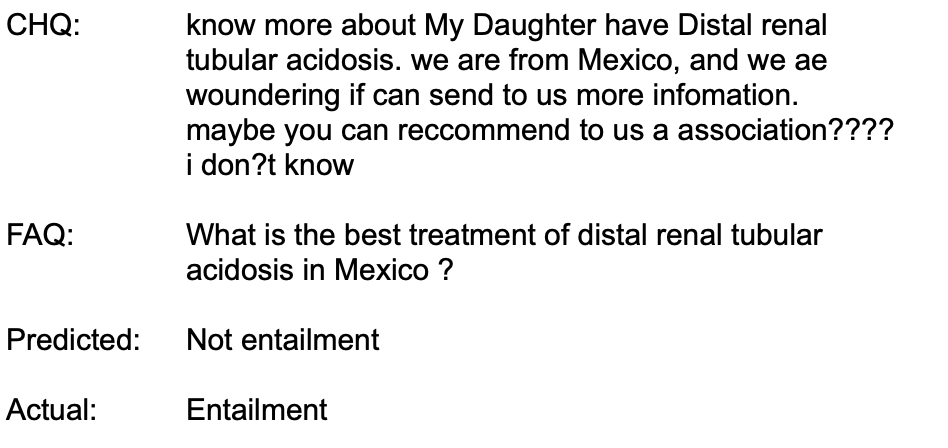}
  \caption{RQE model incorrectly predicting True on test set}
  \label{fig:rqe_error1}
\end{figure}

\begin{figure}[h!]
\centering
  \includegraphics[scale=0.49]{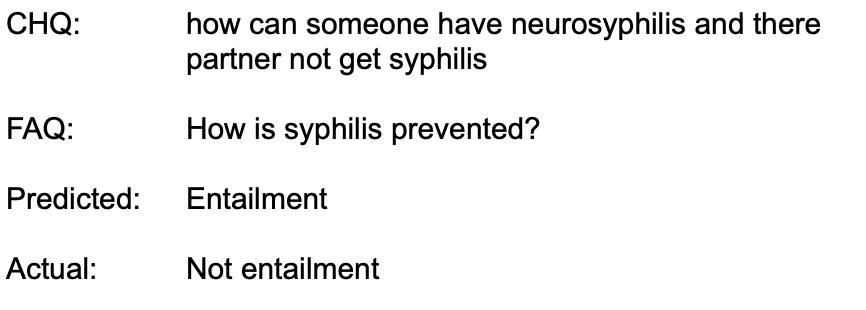}
  \caption{RQE model incorrectly predicting False on test set}
  \label{fig:rqe_error2}
\end{figure}

The example shown in Figure \ref{fig:rqe_error1} has a unigram overlap of 6 and bigram overlap of 3. So our model predicts the label as True, whereas the ground truth label is False because even though the same disease is being referred to in both the CHQ and FAQ, the questions being asked about it are different.

The example shown in Figure \ref{fig:rqe_error2} has a unigram overlap of 2 and bigram overlap of 0. So our model predicts the label as False, whereas the ground truth label is True because the FAQ is sort of like an abstractive summary of the CHQ with less lexical overlap. 

Above analysis suggests that RQE or NLI models are baised to the lexical overlap of medical entities. To overcome this, we could extract medical entities using Metamap \cite{aronson2006metamap} and mask them randomly during training so that the model learns the semantic representation even without the medical entities. Masking entities has been shown to generalize better in ERNIE\cite{2019arXiv190507129Z} in comparison to BERT\cite{devlin2018bert}. 

\begin{figure}[h!]
  \includegraphics[scale=0.42]{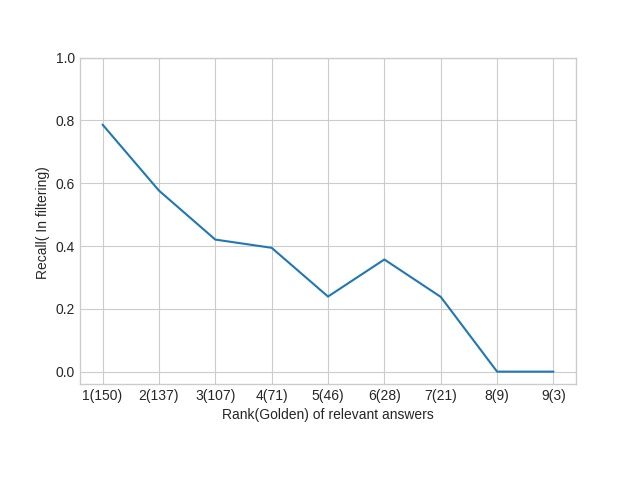}
  \caption{Relationship between the rank of the valid answer and it's filtering recall. The number in parenthesis denotes number of such examples seen in the test dataset.}
  \label{fig:rank_error}
\end{figure}

\begin{figure}[h!]
  \includegraphics[scale=0.42]{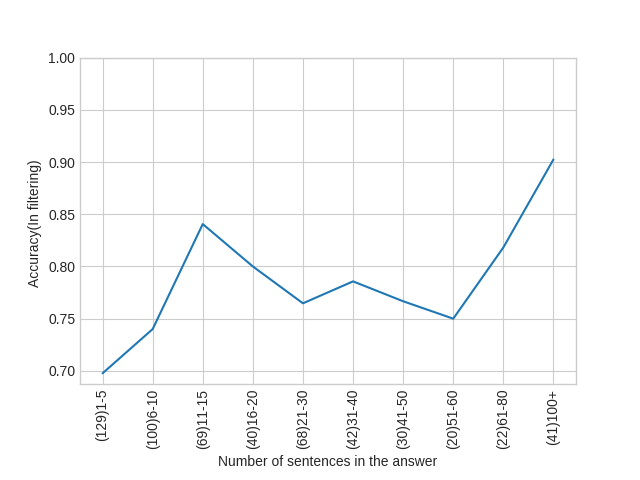}
  \caption{Relationship between the number of sentences in an answer  and the filtering accuracy. The number in parenthesis denotes number of such examples seen in the test dataset.}
  \label{fig:len_error}
\end{figure}

For the re-ranking and filtering tasks we look into the macro-trends and investigate what qualifies as tougher problems for both the tasks. From Figure \ref{fig:rank_error}, it is clear that lower ranked valid answers are generally harder answers for filtering. Observing the valid answers with low ranks, we see that they generally have only 1-2 relevant sentences each, which might be hard for the model especially in cases where the answers have a lot of sentences.   
Similar analysis for the filtering tasks based on the number of sentences in the answers  show some interesting trends, as shown in Figure \ref{fig:len_error}. Interestingly, the model performs really well for filtering longer answers with more than 80 sentences. On further analysis, it is seen that generally the entailed answers can be entirely found in these large candidate answers for the valid answers. 

\begin{figure}[h]
  \includegraphics[scale=0.42]{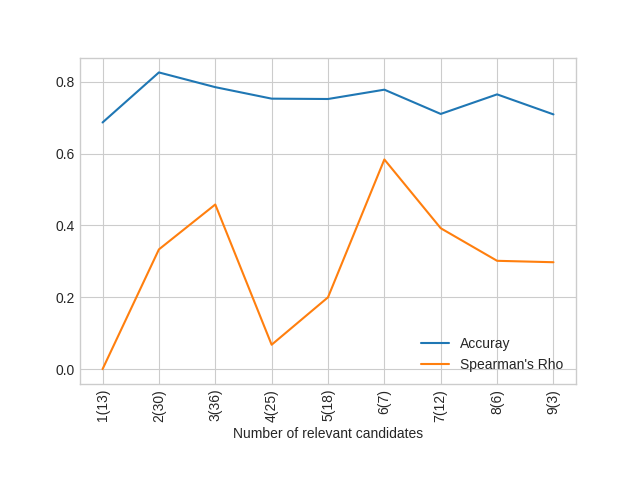}
  \caption{Trends in accuracy and spearman's rho based on the number of valid answers for a question. In case of 1 valid answer, spearman's rho is always taken as 0. The number in parenthesis denotes number of such examples seen in the test dataset.}
  \label{fig:corr_error}
\end{figure}

We also observe that the spearman's rho is sensitive to the number of valid candidates for each question. Especially when the number of valid candidates are less, the metric can vary considerably even with a small error. When analyzing the spearman's rho on per question basis, it is seen that the questions with just two valid answers get a score of -1 on getting the order wrong, while the score is 1 if the order is right. This variability is captured in Figure \ref{fig:corr_error}. The accuracy however, varies only slightly based on the number of valid answers.

\section{Conclusion and Future work}
Our results show that learning to re-rank and filter answers in a multi-task setting help learn a joint feature space which improves performance on both the tasks. In addition, we show that we can harness the power of pre-trained models by fine-tuning them for a specific task and using them as feature extractors to assist in non-trivial tasks such as re-ranking and large document classification.   
    We see that an increase in the size of the corpus with augmented data leads to improved results, hence some more work can be done to build upon the work of \cite{2019arXiv190108079B}. Additionally, we could improve the NLI and RQE systems by tackling the bias created due to the lexical overlap of medical entities among the two sentences/questions, as these were the predominant errors made by our models. This would indirectly translate to an improved performance of the filtering and re-ranking system. 
\bibliography{acl2019}
\bibliographystyle{styles/acl_natbib}
\appendix
\section{Appendix}
\label{sec:appendix}

\begin{table}[h]
    
      
\begin{tabular}{|l|l|}
\hline
Classifier & Architecture \\ \hline
Filtering & \begin{tabular}[c]{@{}l@{}}3824-2048:bn:a\\ 2048-1024:bn:a\\ 1024-512:bn:a\\ 512-512:bn:a\\ 512-256:bn:a\\ 256-64:bn:a\\ 64-1:a\end{tabular} \\ \hline
Pairwise Ranking & \begin{tabular}[c]{@{}l@{}}7648-3824:bn:a\\ 3824-2048:bn:a\\ 2048-1024:bn:a\\ 1024-512:bn:a\\ 512-512:bn:a\\ 512-256:bn:a\\ 256-64:bn:a\\ 64-1:a\end{tabular} \\ \hline
\end{tabular}
\caption{Classifier Specifications: `X1-X2' - denotes a linear layer with X1 input features and X2 output features. `bn’ - with batch normalization, `a': denotes activation, `:' - separates two layers. Activation used everywhere is ReLU except for the output layer where sigmoid is used.}
\label{tab:classifier}
\end{table}
\begin{table}[h]
\begin{tabular}{|l|}
\hline
Convolution Encoder Layers \\ \hline
Input :  c :768 \\ \hline
c:768, k:(1,1), s:(1,1),p:(1,1), bn \\ \hline
c:512, k:(3,3), s:(1,1), p:(2,2), bn \\ \hline
c:512, k:(3,3), s:(2,2), p:(1,1) \\ \hline
c:256, k:(2,2), s:(1,1), p:(1,1), bn \\ \hline
c:,256 k:(3,3), s:(1,1), p:(2,2), \\ \hline
Quadrant Pooling \\ \hline
\end{tabular}
\caption{ Convolution Encoder  Specification.  `c': number of filters, `k': kernel size, `s': stride
size, `p': padding size, `bn': with batch normalization. The sizes are in order (height, width). ReLU activation function is used after each layer except for the input and output layer. Quadrant Pooling is described in Section \ref{sec:approach}.} 
\label{tab:arch}
\end{table}

\end{document}